# A Study of Efficient Energy Management Techniques for Cloud Computing Environment

Syed Arshad Ali, Student Member, IEEE, Mohammad Affan, Mansaf Alam, Member, IEEE
Department of Computer Science
Jamia Millia Islamia
New Delhi, India
arshad158931@st.jmi.ac.in, affan940@gmail.com, malam2@jmi.ac.in

*Abstract*— **The overall performance of the development of computing systems has been engrossed on enhancing demand from the client and enterprise domains. but, the intake of ever-increasing energy for computing systems has commenced to bound in increasing overall performance due to heavy electric payments and carbon dioxide emission. The growth in server's power consumption is increased continuously; and many researchers proposed, if this pattern repeats continuously, then the power consumption cost of a server over its lifespan would be higher than its hardware prices. The power intake troubles more for clusters, grids, and clouds, which encompass numerous thousand heterogeneous servers. Continuous efforts have been done to reduce the electricity intake of these massive-scale infrastructures. To identify the challenges and required future enhancements in the field of efficient energy consumption in Cloud Computing, it is necessary to synthesize and categorize the research and development done so far. In this paper, the authors discuss the reasons and problems associated with huge energy consumption by Cloud datacentres and prepare a taxonomy of huge energy consumption problems and its related solutions. The authors cover all aspects of energy consumption by Cloud datacentres and analyze many research papers to find the better solution for efficient energy consumption. This work gives an overall information regarding energy-consumption problems of Cloud datacentres and energy-efficient solutions for this problem. The paper is concluded with a conversation of future enhancement and development in energy-efficient methods in Cloud Computing.**

## I. Introduction

Cloud Computing is the uprising model in ICT that makes computing trustworthy, dynamic, convenient, and rapidly fast. Cloud Computing has merits over traditional computing as it possesses nimbleness, device undependability and ascendable characteristic [1][2][3]. A Datacenter is like a farmstead that holds huge number of servers which provide data management, networking, data storage, backup and recovery [4]. Half of the energy supplied to a datacenter is used by cooling system at the infrastructure level and a lot of energy is being used by systems when they are in the idle state [5]. Both the providers and users get financial loss from such type of wastes. As per the report by Gartner [6], the cloud datacenters having the electricity consumption, that will be increased to the 012.02 Billion kWh by 2020. By considering the complications related to datacenters, energy efficiency and the trustworthiness are the two major challenges in today's scenario. With the growing capabilities of datacenters, if efficient power management tactics are not applied, then power consumption of datacenters will be increased continuously. The report [7] says, the total worldwide power consumption by the datacentres in 2015 was almost 416.2 terawatt which is more than the energy consumption of the United Kingdom. The Cloud Computing infrastructure is one of the major importers to carbon emission. Cloud vendors like Google, Amazon have been working so as to produce the cloud services that are eco-friendly and efficient [8]. In response to these, several researchers around the world have suggested many algorithms architectures and different policies build the cloud computing environment trustworthy and power efficient. This paper gives the overview and classification of enormous energy consumption problems and its related solutions also it suggests various energy efficiency techniques to improve resource utilization and power consumption in cloud datacenters.

Rest sections of the paper are as follows. In section II, related works have been discussed. Section III gives a taxonomy of various energy efficient techniques and discussion on these techniques are mentioned in section IV. Section V concludes the paper.

## II. Related works

Cloud computing has changed the IT industry by providing an elastic on-demand allocation of computational resources comprising processors, storage & networks which is accompanied by creation, modification, and enhancement of large-scale systems consisting of cluster, grids and Cloud datacenters. This system leads to a lot of energy consumption and significant $CO_2$ emissions. Due to the increased quantity of computational resources, the energy bills come as the second largest items in the budgets of Cloud service providers. Many researchers have been worked to enhance the effective energy utilization in Cloud datacenters and proposed many algorithms related to virtual machine migration, consolidation, and VM allocation. Various task scheduling parameters in Cloud Computing have been discussed in the paper [9], energy-efficiency is more concern parameter of today's research community. In paper [10], the author gives an absolute review of currently alive techniques for energy regulation and reliability. To improve the availability for the cloud services, energy-aware resource provisioning strategy is identified simultaneously minimizing its energy consumption. Several challenges and research gaps for future research and developments for a trade-off between energy regulation and reliability are also identified. Another review paper [11] summed up with a few existing energy scheduling algorithms taken up in a cloud environment also the energy saving percentage in present energy-efficient scheduling algorithms. The results reveal the best energy saving proportion level can be attained by using DVFS and DNS both. In paper [12], the author discussed the double role of Cloud Computing as a huge power consumer and as an energy saving method with compare to traditional computing systems. This paper gives a comprehensive and relative study of several energy efficient methods in Cloud Computing. The power consumption of ICT equipment is discussed in the paper [13]. The authors give a classification of power and system performance based efficient methods for grid, cluster and Cloud datacenters. This survey is different from other surveys because it discussed both aspects of power consumption and system performance of ICT equipment. In this paper, the author also presents a taxonomy of energy-efficient techniques and discussed various methods, concerning energy efficiency and other related parameters of Cloud Computing.

III. TAXONOMY OF ENERGY-EFFICIENT TECHNIQUES

Energy conscious scheduling such as DVFS, energy efficient load balancing, virtualization, resource consolidation, and migration are mostly reviewed for knowledge and practical implementations. Many researchers worked for efficient power consumption in Cloud Computing. In this paper author categories, these techniques in different ways and describes the method, improvements, and limitation of these techniques.

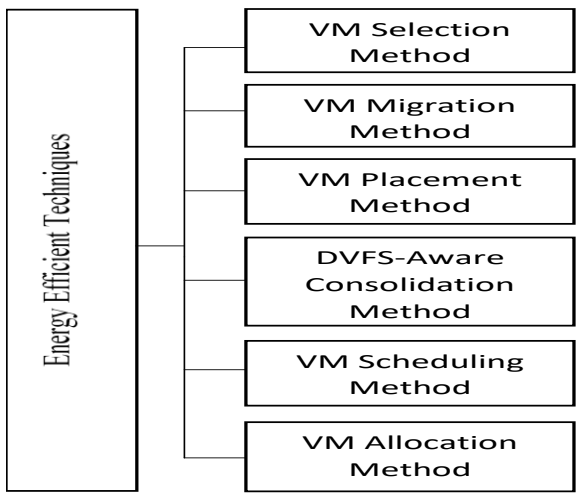

Fig. 1. Taxonomy of Energy-Efficient techniques for Cloud Computing

Figure-1 presents the taxonomy of various energy efficient techniques. The author further discusses all these techniques one by one and give a detailed description and study of these algorithms in table-1.

*A. Virtual Machine (VM) Selection Methods*

- Random VM selection: A uniform distributed discrete random variable is used to select the virtual machine from the overloaded server for migration [14].
- Minimum migration time: Migration time of the virtual machine is considered as the ratio of the quantity of RAM utilization of virtual machine to the server's bandwidth that hosted virtual machine. In this method of VM selection for migration, a VM having minimum migration time is select for migration in comparison to other VMs [15] [16].
- Minimum utilization: Physical resource utilization by the virtual machine is considered as the ratio of the volume of resource utilized by the virtual machine due to user's tasks allocated to that VM and total MIPS allocated to that VM. And in this method [15] a VM which has minimum utilization has to be select for migration.
- Least VM in CPU utilization first: In this method of VM selection [17], a VM which has share least CPU time with other virtual machines (VMs) allocated to the same server has been selected for migration.
- Maximum correlation: Multiple correlation coefficient [18] is used to calculate the correlation among the virtual machines hosted on the same server. In this method [15] [19] a VM with a maximum correlation of CPU usage with the others VMs has been selected for migration.

*B. Virtual Machine (VM) Placement Methods*

For developing Cloud datacentres, energy consumption is the main concern. Several methods and techniques have been proposed to reduce energy consumption but these techniques are mainly having more VM migration and less resource utilization.

In paper [20] the authors proposed a VM placement technique based on a heuristic greedy algorithm. In this algorithm, the author develops a VM deployment and live migration model to improve resource utilization and decrease power consumption. The heuristic algorithm predicts the workload and mapped CPU-intensive workload and memory-intensive workload to the same physical server to reduce energy usage by the different servers and balancing the workload.

In paper [21], the author developed a statistical mathematical framework for VM placement, which integrates complete virtualization expenses in the dynamic migration process. The proposed dynamic virtual machine placement method enables VM request scheduling and lives migration to reduce the active server participation so as to reduce the power consumptions in Cloud datacentres.

The author proposed a VM allocation method [22], based on minimum virtual machine migration. Three strategies named fixed double threshold, double resource threshold and dynamic double threshold are developed and two phases in each strategy are used. In the first phase, VM selection is made and selected VM is placed on a physical datacentre in the second phase. The result shows that these lower and upper bound resource utilization threshold policies are better than the single threshold technique. These methods determine less power consumption, a smaller number of SLA violations and a minimum number of VM migration.

The main concern of Cloud service provider is to answer two questions, where to place VMs initially and where to transfer VMs, when VM-movements are required. VM migration is helpful to reduce datacentre overloading and reduce active servers' involvement for effective resource utilization and power saving. It is important to detect overloaded servers efficiently for better performance and minimum service cost of Cloud system. In [23], a logistic regression and median absolute derivation methods are used to proposed a general detection algorithm for the overloaded server. Any VM placement and migration algorithms can detect overloaded server with this detection algorithm.

In [24], the authors have presented an algorithm which is task-based virtual machine placement algorithm, in which tasks are mapped to the VMs according to their demand and VMs are placed to physical machine accordingly. The algorithm reduces the number of active servers involved in serving VMs to reduce energy consumption. It also reduces the task rejection rate and makespan of the Cloud system.

### C. Virtual Machine (VM) Migration Methods

An energy consumption model is proposed in paper [25] which is based on the statistical method and can estimate the VM power consumption with the error rate of 3%-6%. In this method, a workload threshold is set for each server, and if a server exceeds its workload threshold then the VM will be migrated from that overloaded server to another server to reduce the energy consumption by the overloaded server. This method can achieve an effective reduction in power consumption without violating the QoS.

A linear integer programming model and bin packaging model are used in [26], to develop two exact algorithms for VMs placement and consolidation for reducing power consumption and VM migration cost and compared with the heuristic based best-fit algorithm. The results show that the combination of these two algorithms contributes to a significant reduction in power consumption.

In the paper [22], the authors have proposed three policies for VM placement and migration. When the number of VM placement increased o server then due to overload the VM migration is required. Which server has to be select for VM migration is dependent on these policies named FDT, DRT, and DDT. These are three different methods for selection of the server for migration according to the threshold set by these policies.

### D. DVFS-Aware Consolidation Methods

More than 43 million ton of $Co_2$ emission per year and about 2% of the world's power production has been consumed by the Cloud datacentres. In paper [27], the author proposed two methods, one for efficient power consumption based on DVFS technique and second for VM consolidation. The first method is used to determine performance degradation with power consumption and gives a DVFS-aware workload management which saves energy up to 39.14% for dynamic workload situations. The second VM consolidation method is also determined dynamic frequency while allocating workload to achieve QoS.

There are different types of physical machines are available in Cloud datacentres. This machine heterogeneity consumes more energy when workloads have been scheduled on them. A job consolidation algorithm with DVFS technique is proposed in paper [28], for efficient resource utilization in heterogenetic Cloud physical machines. The proposed algorithm will replace jobs efficiently to reduce energy consumption.

### E. Virtual Machine (VM) Scheduling Methods

In this paper [29], the author proposed an online scheduling algorithm for IaaS Cloud model for reduction in energy consumption. The algorithm works for heterogeneous machines and different workload scenario to achieve a better quality of service.

One way to reduce energy consumption in Cloud datacentres is to shut down physical servers which are idle. In paper [30], an energy-aware virtual machine scheduling algorithm has been proposed named as dynamic round robin algorithm. The results showed that the algorithm saves 43.7% energy and 60% of physical machine usage compared with other scheduling algorithms.

The authors suggest a model [31] for energy consumption estimation, which considered the running tasks created by virtual machine for estimation of each VM's power consumption. The suggested model also schedules the VMs to confirm the energy cost of each VM.

Most of the energy efficient methods use VM migration technique but in the paper [32], the author proposed an energy-aware virtual machine scheduling algorithm EMinTRE-LFT, which is based on the concept i.e., decrease in power consumption is directly equivalent to minimization in the completion time of all physical servers. The author used OpenStack Nova scheduler for simulation and compare it with other algorithms.

The Cloud scheduling algorithms face many challenges due to the dynamic and unpredictable nature of Cloud user's request. In this paper [33], the author proposed an algorithm which does not require any prior knowledge of user's request. The author conducted a mathematical analysis to find the balance between energy consumption and system performance.

A real-time dynamic scheduling algorithm is proposed in paper [34], which schedule distributed application in a distributed system to reduce the power consumption. The proposed algorithm uses heuristics and resource allocation techniques to get the optimal solution. It minimizes the power consumption and task execution time with order dependent setup between tasks for VM and power setup for different Cloud designs.

### F. Virtual Machine (VM) Allocation Methods

The authors proposed an interior search based virtual machine allocation algorithm for efficient energy consumption and proper resource utilization in the paper [35]. The model and simulation of the proposed algorithm are tested on CloudSim and compared the amount of energy consumption with the Genetic Algorithm (GA) and Best-fit Decreasing (BFD) algorithm.

Cloud provider allocates VMs to the customer's application according to their demand, and these VMs are assigned to the physical machines. Many resource allocation methods use VMs resource utilization history for efficient resource allocation. In paper [36], the author proposed a QoS-aware virtual machine allocation method based on resource utilization history to improve the level of quality of services and reduce energy consumption.

Cloud datacentres provide services to Cloud applications which consume a huge amount of energy and produce carbon emission. To overcome from this issue, the author proposed an energy-aware VM allocation algorithm in [14], that provision and schedule Cloud datacentre resources to the user's tasks in an efficient manner that reduces energy consumption level of datacentres and improve the quality of service.

Many researchers worked for energy efficiency in Cloud Computing but some researchers are working for energy efficiency in a specific type of Datacentres. In paper [37], the author proposed an efficient power consumption algorithm for video streaming datacentres. They proposed a method for VM management with the power-law feature. It predicts the future resource usage of VM, according to the popularity of video and arranges sufficient resources for that VM and shut down the idle servers on the datacentres to reduce power consumption. The results showed that this algorithm reduced more power consumption compared with Nash and Best-fit algorithm.

TABLE I.    STUDY OF VARIOUS ENERGY EFFICIENT ALGORITHMS

| Energy Efficient Algorithm | Techniques | Findings | Parameters Improved | Experimental Tool | Environment |
|---|---|---|---|---|---|
| Energy-aware resource allocation heuristic algorithm [14] | VM-Selection VM-Allocation | Uniform Distributed Discrete Random Variable is used for VM-Selection | Maintain QoS and reduce power consumption | CloudSim toolkit | Cloud |
| Optimal online deterministic algorithm [15] | VM-Selection | For VM-selection minimum migration time of VM has been used | Reduce energy consumption and high level of observance to SLA | CloudSim | Cloud |
| Energy efficient resource utilization algorithm [17] | VM-Selection | Least CPU utilization first VM is selected | Reduce $Co_2$ emission and energy consumption, resource utilization | CloudSim toolkit | IaaS Cloud Model |
| Energy-aware resource allocation algorithm [20] | VM-Placement | The heuristic greedy algorithm is used for workload prediction | Reduce SLA Violation, Energy wastage, Balance Workload and provide Scalability | PowerEdge blade servers are used for experiments | Cloud |
| Dynamic virtual machine placement algorithm [21] | VM-Placement | The statistical mathematical framework is used | Workload handling, reduce power consumption | Self-designed Simulator | Cloud |
| Efficient allocation of VMs in servers [22] | VM-Placement VM-Migration | Three VM-Placement policies FDT, DRT, and DDT are used | Less power consumption, a smaller number of SLA violations and reduce VM migration | CloudSim Simulator | Cloud |
| Energy efficient dynamic resource management [23] | VM-Placement | Logistic regression and median absolute derivation models are used for detection algorithms. | Minimize power consumption of the datacentre and avoid SLA violations. | CloudSim | Cloud |
| Energy-Efficient VM-Placement [24] | VM-Placement | Demand-based VM and Physical machine mapping algorithm is developed | Reduce energy consumption, makespan, and task rejection rate | CloudSim | Cloud |
| An Energy-Saving Virtual-Machine Migration [25] | VM-Migration | Statistical based energy consumption algorithm | Efficient power consumption and provides QoS | PowerEdge blade servers are used for experiments | Cloud |
| Exact allocation and migration algorithms | VM-Migration | Linear integer programming model | Minimize energy consumption and | A dedicated simulator is developed based on | Cloud |

| [26] | | and bin packaging model are used | number of VM migration | Java language implementation and the linear solver CPLEX | |
|---|---|---|---|---|---|
| DVFS-Aware consolidation for energy-efficient clouds [27] | DVFS-Aware VM-Consolidation | Two methods are developed, one for efficient power consumption based on DVFS technique and second for VM consolidation | Energy efficient, maintain QoS and improve system performance. Save energy up to 39.14% for dynamic workload situations | CloudSim | Cloud |
| Towards energy-aware job consolidation scheduling [28] | DVFS-Aware VM-Consolidation | Job consolidation algorithm with DVFS technique | Reduce energy consumption, supports physical machine heterogeneity | CloudSim | Cloud |
| Energy-aware online scheduling [29] | VM-Scheduling | Online VM Scheduling algorithm is developed | Maintain QoS, reduce energy consumption, supports a heterogeneous environment | Self-designed Simulator | Cloud |
| Energy-aware virtual machine dynamic provision and scheduling [30] | VM-Scheduling | Dynamic round robin algorithm is proposed for power-aware VM scheduling. | Reduce energy consumption and physical machine usage | Eucalyptus: An open-source Cloud middleware | Cloud |
| Energy-based accounting and scheduling of virtual machines [31] | VM-Scheduling | In-processor tasks generation model for energy efficiency | Reduce energy consumption, energy-credit scheduler | Xen Hypervisor | Cloud |
| Energy-efficient Scheduling of Virtual Machines in IaaS clouds [32] | VM-Scheduling | Proposed an energy-aware virtual machine scheduling algorithm EMinTRE-LFT | Reduce completion time and power consumption of physical machine | OpenStack Nova Scheduler | Cloud |
| Energy efficient scheduling and management for large-scale services computing systems [33] | VM-Scheduling | Lyapunov optimization technique based distributed the online algorithm | Energy efficiency, prior knowledge of user's request is not required, improved system performance | Numerical experiments and real trace-based simulation | Cloud |
| Dynamic energy-aware scheduling for parallel task-based application [34] | VM-Scheduling | Multi-heuristic resource allocation (MHRA) is proposed | Minimize energy consumption and task execution time | Implement a Scheduler using COMPS programming model | Cloud |
| Energy-Efficient virtual machine allocation technique using interior search algorithm [35] | VM-Allocation | An interior search based virtual machine allocation algorithm | Reduce energy consumption, and efficient resource utilization | CloudSim | Cloud |
| Novel resource allocation algorithms to performance and energy efficiency [36] | VM-Allocation | Proposed UMC and VDT algorithms | Efficient energy consumption and resource utilization, maintain QoS | CloudSim | Cloud |
| An energy efficient VM management scheme with power-law characteristic [37] | VM-Allocation | Proposed a power consumption algorithm for video streaming datacentres | Energy efficiency, efficient resource utilization, video streaming datacentres | CloudSim with Python (pyCloudSim) | Cloud |

## IV. DISCUSSION

Increment in power utilization is the emerging problem in today's computing world. Hike of applications related to complicated data have introduced the establishment of big datacenters which raised the energy need. From the above study of energy efficient techniques, we can say that, most of the work to reduce energy consumption in datacenters are done using VM-migration and VM-scheduling methods. Some researchers proposed multi-objective algorithms, which are mostly cover SLA, QoS and resource utilization with efficient energy consumption in Cloud datacenters. Less work has been done for heterogeneous physical machines, which needs some attention from research community.

## V. CONCLUSION

In this paper, the author presents a taxonomy of energy efficient techniques for Cloud Computing. Various algorithms have been studied and their finding and improved parameters are listed in the table. This paper can help readers to find merits and limitations of proposed energy efficient algorithms present in the literature.

ACKNOWLEDGMENT

This work was supported by a grant from "Young Faculty Research Fellowship" under Visvesvaraya Ph.D. Scheme for Electronics and IT, Department of Electronics & Information Technology (DeitY), Ministry of Communications & IT, Government of India.